\documentclass[preprint2]{aastex6}

\newcommand{\source}{4U$\,$0142+61}

\newcommand{\myemail}{rarchiba@physics.mcgill.ca}

\shorttitle{{\it Swift} Monitoring of \source{}}
\shortauthors{Archibald et al.}
\usepackage{threeparttable}
\usepackage{lineno, soul, color}

\begin{document}

\title{{\it Swift} Observations of Two Outbursts from the magnetar \source{}}

\author{R. F. Archibald\altaffilmark{1}$^{\dagger}$, V. M. Kaspi\altaffilmark{1},  P. Scholz\altaffilmark{1,2}, A. P. Beardmore\altaffilmark{3}, N. Gehrels\altaffilmark{4}, \& J. A. Kennea\altaffilmark{5}}

 \altaffiltext{1}{Department of Physics and McGill Space Institute, McGill University, 3600 University Street, Montreal QC, H3A 2T8, Canada\\$^\dagger$E-mail: \myemail}
 \altaffiltext{2}{National Research Council of Canada, Herzberg Astronomy and Astrophysics, Dominion Radio Astrophysical Observatory, P.O. Box 248, Penticton, BC V2A 6J9, Canada}
 \altaffiltext{3}{Department of Physics and Astronomy, University of Leicester, University Road, Leicester LE1 7RH, UK}
 \altaffiltext{4}{Astrophysics Science Division, NASA Goddard Space Flight Center, Greenbelt, MD 20771 USA}
 \altaffiltext{5}{Department of Astronomy and Astrophysics, 525 Davey Lab, Pennsylvania State University, University Park, PA 16802, USA}

\begin{abstract}
\source{} is one of a small class of persistently bright magnetars. Here we report on a monitoring campaign of \source{} from 2011 July 26 - 2016 June 12 using the {\it Swift} X-ray Telescope, continuing a 16 year timing campaign with the {\it Rossi X-ray Timing Explorer}.
We show that \source{} had two radiatively loud timing events, on 2011 July 29 and 2015 February 28, both with short soft $\gamma$-ray bursts, and a long-lived flux decay associated with each case.
We show that the 2015 timing event resulted in a net spin-down of the pulsar due to  over-recovery of a glitch. We compare this timing event to previous such events in other high-magnetic-field pulsars, and discuss net spin-down glitches now seen in several young, high-B pulsars.
\end{abstract}

\section{Introduction}

Magnetars are a class of young, rotating neutron stars with spin periods ranging from $\sim$0.3--12$\;$s.
They appear to have the highest spin-down-inferred dipolar surface magnetic fields of the neutron-star population, with  magnetic fields of order $10^{14}-10^{15}\;$G, although recently magnetars with more typical inferred dipole magnetic fields have been found \citep[e.g.][]{2013ApJ...770...65R, 2014ApJ...786...62S}.
Magnetars exhibit a wide array of radiative behavior,  showing flux variability at nearly all timescales -- from soft-$\gamma$-ray bursts lasting milliseconds, to orders of magnitude X-ray flux increases which decay over months to years.
As implied by the name, it is the decay of their magnetic fields that is thought to power the X-ray emission of these objects, the energy of which can be much greater than that available via the spin-down luminosity which accounts for rotation-powered pulsar energetics \citep{1995MNRAS.275..255T, 1996ApJ...473..322T}.
For a recent review, see \cite{2015SSRv..191..315M}.
There are currently 23 confirmed magnetars; an up-to-date list of both confirmed and candidate magnetars is maintained in the McGill Magnetar Catalog\footnote{\url{www.physics.mcgill.ca/~pulsar/magnetar/main.html}} \citep{2014ApJS..212....6O}.

Magnetars, like rotation powered pulsars, experience sudden changes in their spin-frequency known as glitches \cite[see e.g.][]{2000ApJ...537L..31K, 2008APJ.673..1044D}.
While similar in many regards to glitches in rotation-powered pulsars, magnetar glitches are unique in that they are often accompanied by radiative changes including short soft-$\gamma$-ray bursts, pulse profile changes, and long-term (months to years) X-ray flux enhancements  \citep[see][for a review]{2011ASSP...21..247R, 2014ApJ...784...37D}.

\source{} has the highest persistent X-ray flux of the known magnetars.
It was discovered as a point-source by the {\it Uhuru} mission \citep{1972ApJ...178..281G}, and later revealed to be a $\sim$8.7$\;$s pulsar \citep{1994ApJ...433L..25I}.
It has an inferred dipole  surface magnetic field of $1.3\times10^{14}$\,G.

\source{}, along with several other magnetars,  was monitored regularly with the {\it Rossi X-ray Timing Explorer (RXTE)} in 1997, and from 2000 until the decommissioning of {\it RXTE} in 2011 December \citep{2014ApJ...784...37D}.
During this campaign \source{} showed a slow rise in pulsed flux accompanied by pulse profile changes from 2000--2006 \citep{2007ApJ...666...1152D}.
Most notably, in 2006, \source{} entered an active phase -- emitting at least six short magnetar-like bursts, and exhibiting a glitch with an over-recovery, leading to a net spin-down of the neutron star \citep{2011ApJ...736..138G}.
On 2011 July 29, \source{} had large spin-up glitch \citep{2014ApJ...784...37D}.
This glitch was accompanied by a short hard X-ray burst detected by the  {\it Swift} Burst Alert Telescope (BAT) \citep{2011ATel.3520....1G}.

Compared to the other long-term {\it RXTE}-monitored magnetars, \source\ is relatively quiet in terms of timing behavior.
Aside from the aforementioned glitches, the rotational evolution is well described by a low-order polynomial, a statement which cannot be made for any of the other sources in the {\it RXTE} monitoring campaign \citep{2014ApJ...784...37D}.

Here we present the results of a continued monitoring campaign of \source{} using the {\it Swift} X-ray Telescope (XRT) from 2011 July to 2016 June.
We report on two X-ray outbursts on 2011 July 29 and 2015 February 28, both associated with timing anomalies.
 We find that the timing anomaly associated with the X-ray outburst on 2015 February 28, similar to the 2006 event, led to a net spin-down of \source{}.
We also find a long-term X-ray flux decay following both timing events, as well as 12  XRT-detected magnetar-like bursts which occurred coincident with the second timing event.

\section {Observations and Analysis}

\subsection{{\it Swift} X-ray Telescope}
\label{sec:xrt}
We began observing \source{} with the {\it Swift} XRT on  2011 July 26 as part of a campaign to monitor several magnetars \citep[see e.g.][]{2014ApJ...783...99S, 2015ApJ...800...33A}.

The {\it Swift} XRT \citep{2005SSRv..120..165B} is a Wolter-I telescope with a e2v CCD22 detector, sensitive in the 0.3--10-keV range.
The XRT was operated in Windowed-Timing (WT) mode for all observations,
having a time resolution of $1.76\;$ms, at the expense of one dimension of spatial resolution.

Level 1 data products were obtained from the HEASARC \emph{Swift} archive, reduced using the  {\tt xrtpipeline} standard reduction script, and time corrected to the Solar System barycenter using the position of \source{} \citep{2004A&A...416.1037H}, and {\tt HEASOFT v6.17}.
Individual exposure maps, spectrum, and ancillary response files were created for each orbit and then summed.
If, in an individual orbit, the center of the source was within three pixels of a dead column, or the edge of the chip, that orbit was excluded from flux and spectral fitting.
We selected only Grade 0 events for spectral fitting as other event Grades are more likely to be caused by background events \citep{2005SSRv..120..165B}.
We also removed many detected soft X-ray bursts which appear in both the source and background region, as these bursts must be instrumental in origin.

To investigate the flux and spectral evolution of \source{}, a 10-pixel radius circle centered on the source was extracted.
As well, an annulus of inner radius 75 and outer radius 125 pixels centered on the source was used to extract background events.

{\it Swift} XRT spectra were extracted from the selected regions using {\tt extractor}, and fit using {\tt XSPEC} version 12.8.2\footnote{\url{http://xspec.gfsc.nasa.gov}}.
Photons were grouped to ensure one photon was in each spectral bin.
As the background dominates the source below 0.7 keV, we use photons from only the 0.7--10-keV band for spectral fitting.
In total, 127 XRT observations totaling 475 ks of observing time were analyzed in this work.

\section{Timing Analysis}
\label{sec:timing}

Following the processing described in \S\ref{sec:xrt}, we derived an average pulse time-of-arrival (TOA) for each observation.
To maximize the signal-to-noise ratio for obtaining TOAs for the source, photons from 0.7-10$\;$keV were used in the timing analysis.
The TOAs were obtained using a Maximum Likelihood (ML) method as described in  \cite{2009LivingstoneTiming} and \cite{2012ApJ...761...66S}.
The ML method compares a continuous model of the pulse profile to the photon arrival phases obtained by folding a single observation.

These TOAs were fitted to a standard pulsar timing model wherein the phase, $\phi$, as a function of time, $t$, is described by a Taylor expansion:
\begin{equation}
\phi(t) = \phi_0+\nu_0(t-t_0)+\frac{1}{2}\dot{\nu_0}(t-t_0)^2+\frac{1}{6}\ddot{\nu_0}(t-t_0)^3+\cdots
\end{equation}
where $\nu_0$ is the rotational frequency of the pulsar at $t_0$, $\dot{\nu_0}$ the spin-down rate at  $t_0$, and  $\ddot{\nu_0}$ the second time derivative of the rotational frequency.
This was done using the TEMPO2 \citep{2006MNRAS.369..655H} pulsar timing software package.

As we have only one {\it Swift} observation before the 2011 glitch on MJD 55771.19 \citep{2014ApJ...784...37D} 
we present a timing solution starting at MJD 55771.9, the first XRT observation following the aforementioned glitch.
We note that the timing solution presented by \cite{2014ApJ...784...37D} accurately describes our measured TOAs in the overlapping region.

\begin{table}
\begin{center}
\caption {Phase-Coherent Timing Parameters for \source{}.}
\label{tab:timing}
\begin{tabular}{ll}
\hline
\hline
Dates (MJD)         & 55771.54 - 57551.22 \\
Dates               & 29 July 2011 - 12 June 2016 \\
Epoch (MJD)         & 57000.00000\\
$\nu\;$ (s$^{-1}$)            & 0.115 085 312 4(3)\\
$\dot{\nu}\;$ (s$^{-2}$)            & $-$2.621(2)$\times 10^{-14}$\\
$\ddot{\nu}\;$ (s$^{-3}$)            & $8(3)\times 10^{-25}$ \\
\multicolumn{2}{c}{Glitch Parameters}\\
Glitch Epoch  &  57081.21605 (fixed)\\
$\Delta\nu\;$ (s$^{-1}$)            & $-3.7(1)\times10^{-8}$\\
$\Delta\nu_d\;$ (s$^{-1}$)            & $5.1(5)\times10^{-8}$\\
$\tau_d\;$ (days)            & $57(9)$\\
rms residual (s) & 0.118\\
rms residual (phase) & 0.014\\

\hline
\hline
\end{tabular}
Figures in parentheses are  the nominal 1$\sigma$ \textsc{tempo2} uncertainties in the least-significant digits quoted.
\end{center}
\end{table}

We present a fully phase-coherent timing solution in Table~\ref{tab:timing}. 
The basic timing solution provides an accurate description of the TOAs until MJD  57079.7.
At this date we require a change in the spin parameters to accurately describe the TOAs.
We fit for a glitch with both permanent and decaying parameters wherein the spin frequency after the glitch epoch, $t_g$, can be described as:
\begin{equation}
    \nu(t) = \nu_t+\Delta\nu+\Delta\nu_{d}e^{-(t-t_g)/\tau_d}
\end{equation}
where $\nu_t$ is the predicted spin frequency  pre-glitch, $\Delta\nu$ is a permanent change in the spin frequency, and $\Delta\nu_{d}$ is an exponentially decaying change in the spin frequency decaying with a timescale of $\tau_d$ days.
This glitch is coincident with the 2015 February 28 BAT detection, and short-term X-ray flux increase \citep[see][]{2015GCN..17507...1B} and \S~\ref{sec:shortflux}.

\begin{figure}
\centering
\includegraphics[width=\columnwidth]{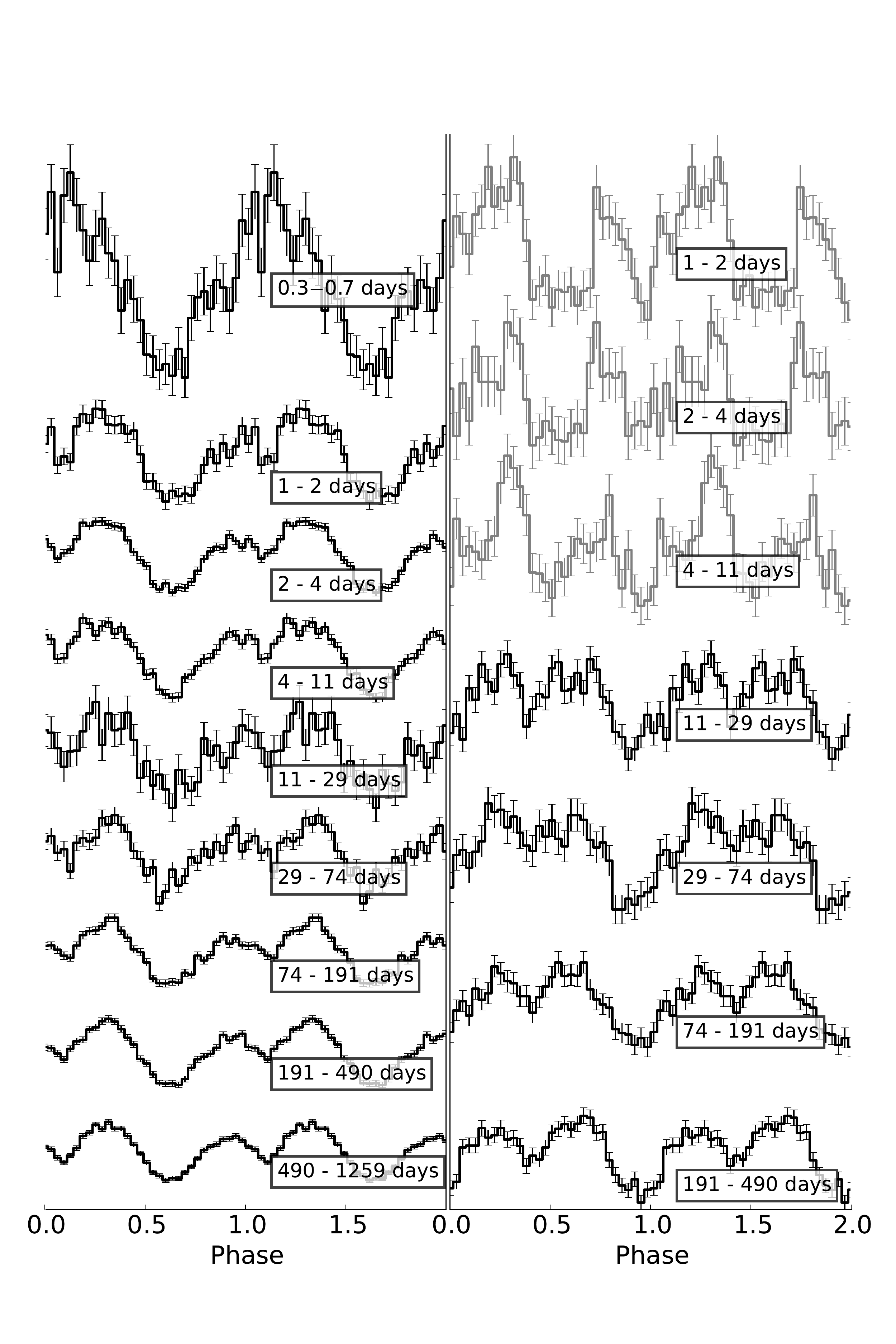}
\caption {Pulse shape evolution of \source{}. The  panels show the total 0.7--10-keV profiles, binned in equal log-time bins. The left panel present profiles following the 2011 outburst, in days from MJDs 55771.19, and the right panel  following the 2015 outburst, in days from MJD 57081.2. The profiles have been normalized, and vertically shifted to avoid overlap. The three profiles in gray are those excluded from the timing solution due to profile variations.}
\label{fig:profs}
\end{figure}

\begin{figure}
\centering
\includegraphics[width=\columnwidth]{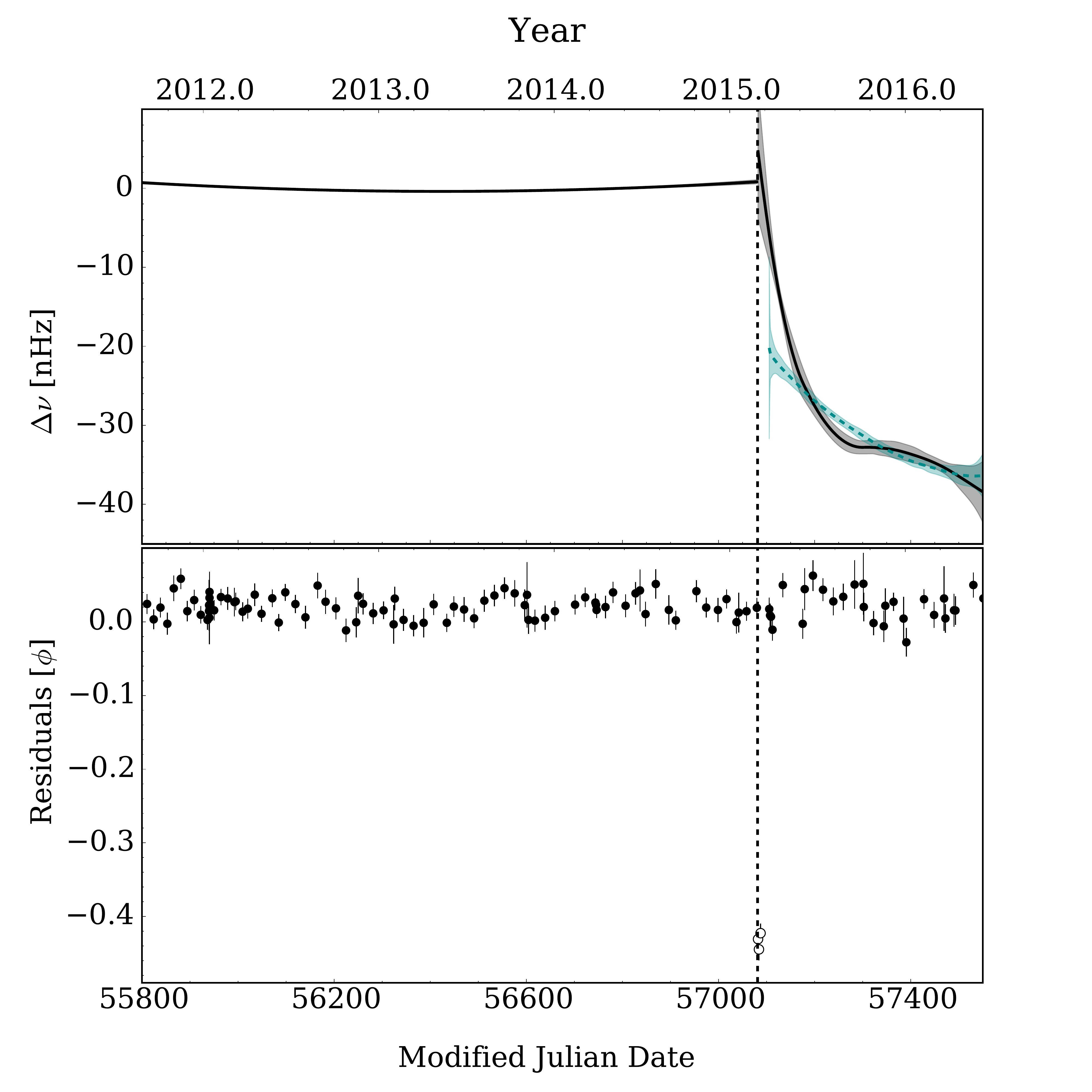}
\caption{Timing evolution of \source{} over the {\it Swift} campaign. The top panel shows $\nu$ over time with the average pre-burst $\dot{\nu}$ trend removed. The solid black line with a gray error region uses all TOAs, and the blue dashed line shows the solution excluding TOAs with strong pulse profile changes; see \S\ref{sec:timing} for details. The bottom panel shows the timing residuals for the ephemeris presented in Table~\ref{tab:timing}. The hollow points denote the TOAs excluded due to profile variations. In both panels, the dashed vertical time indicates the time of the glitch.}
\label{fig:timing}
\end{figure}

Due to pulse shape variations,  for the three observations immediately  following the 2015 outburst --those occurring on MJDs 55782.6, 57084.1, and 57087.5, our ML TOA extraction method indicated that these 3 TOAs were $0.43$ phase turns out of phase with all the surrounding TOAs.
The profiles can be seen in Figure~\ref{fig:profs}.
This offset is consistent with the distance between the two peaks in the standard profile.
We therefore exclude these three TOAs from our timing analysis.
The full timing solution is presented in Table~\ref{tab:timing}.
The timing residuals, the difference between the modeled and observed TOAs, can be seen in Figure~\ref{fig:timing}.
We note that these three suspect TOAs are in phase with each other, and that if we shift them by the time between the two profile peaks, and include them in our glitch fitting procedure, they have little effect ($\sim1\sigma$) on the reported parameters.

We emphasize that the net spin-\emph{down} glitch properties are most affected by the TOAs long after the outburst, far away from the period of strong profile changes.
To illustrate the difference in the inclusion of the suspect profiles, in Figure~\ref{fig:timing} we present the evolution of $\nu$  over the {\it Swift} campaign.
To generate this Figure, we fit splines to the pulse numbers \citep[see][]{splineref}, using a method similar to that described in \cite{2014ApJ...784...37D} using piecewise polynomials of degree $n=3$ weighted by the inverse square error on the pulse number.
To generate the error bounds, we added Gaussian noise with standard deviations of the measured TOA uncertainties to these pulse numbers 1000 times, and refit the splines.
The plotted error band shows the 68\% confidence region.
Note that while in this semi-coherent solution a spin-up glitch is not required when excluding the three suspect TOAs, a fully coherent solution requires the spin-up component in either case.

\section{Radiative properties}
\subsection{Long-term Spectral Evolution}
\label{sec:longflux}
Following the data reduction described in \S~\ref{sec:xrt} we fitted each observation using the typical phenomenological two-component model used for magnetar spectra -- an absorbed blackbody plus a power law.
Photoelectric absorption was modeled using {\tt XSPEC} {\tt tbabs} with abundances from  \cite{2000ApJ...542..914W}, and photoelectric cross-sections from \cite{1996ApJ...465..487V}.

To determine a self-consistent  $N_H$ for \source{}, we simultaneously fit  observations between MJD 56260--56999, i.e. a large set of observations far away from hard X-ray burst detections, and where the flux and spectral parameters of the individual observations were consistent.
We fit these to a single absorbed blackbody plus power law and obtained $N_H=(1.11\pm0.04)\times10^{22}\;$cm$^{-2}$ at 90\% confidence with a C-statistic \citep{1979ApJ...228..939C} of 18779.89 for 18553 degrees of freedom.
As such, for all other spectral results presented here, we have assumed a constant value of $N_H=1.11\times10^{22}\;$cm$^{-2}$.

\begin{figure}
\centering
\includegraphics[width=\columnwidth]{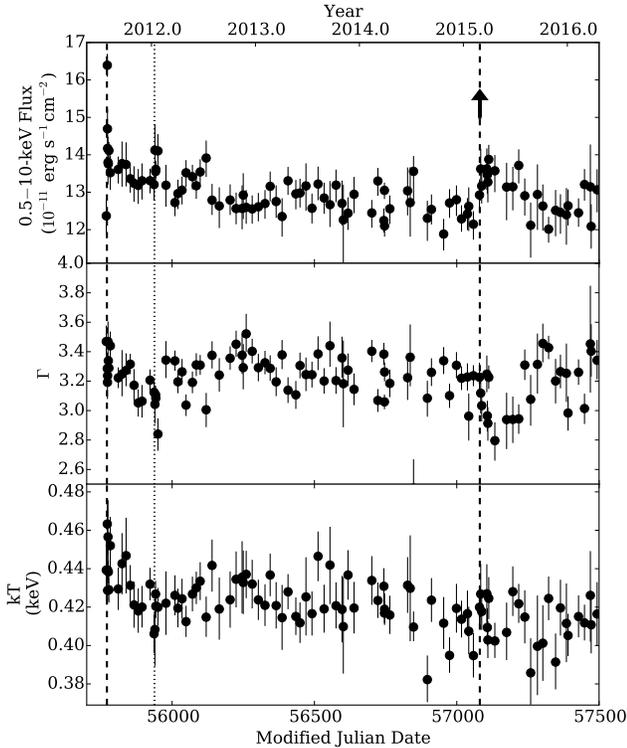}
\caption{Flux and spectral evolution of \source{} during {\it Swift} monitoring. The top panel shows the absorbed 0.5--10-keV flux. The vertical dashed lines indicate {\it Swift} BAT triggers associated with glitches, and the vertical dotted line indicates the BAT trigger with no associated timing behavior. The black arrow indicates the point at which the flux increases by a factor of ten, and decays on minutes timescales (see \S\ref{sec:shortflux} and Fig.~\ref{fig:lc} ).
 The central panel shows $\Gamma$, the power-law index and the bottom panel shows $kT$, the blackbody temperature as a function of time.}
\label{fig:flux}
\end{figure}

In Figure~\ref{fig:flux} we show the absorbed 0.5--10-keV flux, the power-law index, $\Gamma$, and the blackbody temperature, $kT$ for each observation.
We note that these two parameters are highly covariant due to their similar contributions to the flux in this band, and urge caution in interpreting any apparent trend.
The epochs where \source{} triggered the {\it Swift} BAT are indicated on the plot with vertical black lines: the dotted black line indicates the BAT trigger without a corresponding glitch, and the dashed lines indicate those assocated with glitches.
As well, the black arrow on the plot indicates that on MJD 57081.2 the flux was several times higher than the persistent level, decaying within a single orbit and contained several magnetar-like bursts. See \S~\ref{sec:shortflux} for further details.

Taken with the marginal pulsed flux increase seen with {\it RXTE} \citep{2014ApJ...784...37D} we can confirm using the XRT data that the 2011 glitch was a radiatively loud event.
To characterize this radiative behavior, we tried fitting the inter-glitch flux decay between MJDs 55771 and 57079 with a power-law, a linear decay, and an exponential decay, all plus a quiescent flux.
As none of these single component models provided a statistically acceptable fit, we then fitted models with either two power-law decays ($\chi^2/dof=90.8/70$), or two exponentials, both plus a constant quiescent flux.
The best-fitting of these models, with $\chi^2/dof=48.0/70$ is the double exponential, as described by the following equation:
$F(t)=F_Q+F_1e^{-(t-t_0)/\tau_1}+F_2e^{-(t-t_0)/\tau_2}$
where $F_Q$ is the `quiescent' flux, fixed at $12.4(4)\times 10^{-11}$ erg s$^{-1}$ cm$^{-2}$, the level measured on MJD 55768.7, the date of the last XRT observation prior to the flux increase.
The best-fit parameters are $F_1=13(7)\times10^{-11}$ erg s$^{-1}$ cm$^{-2}$, $\tau_{1} = 0.6(2)\;$ days, $F_2=1.4(1)\times 10^{-11}$ erg s$^{-1}$ cm$^{-2}$, and $\tau_2=510(80)\;$ days.
Note that all values here are absorbed 0.5--10-keV fluxes, and $t_0$ is held fixed at the time of the {\it RXTE} reported glitch, MJD 55771.19.

\begin{figure}
\centering
\includegraphics[width=\columnwidth]{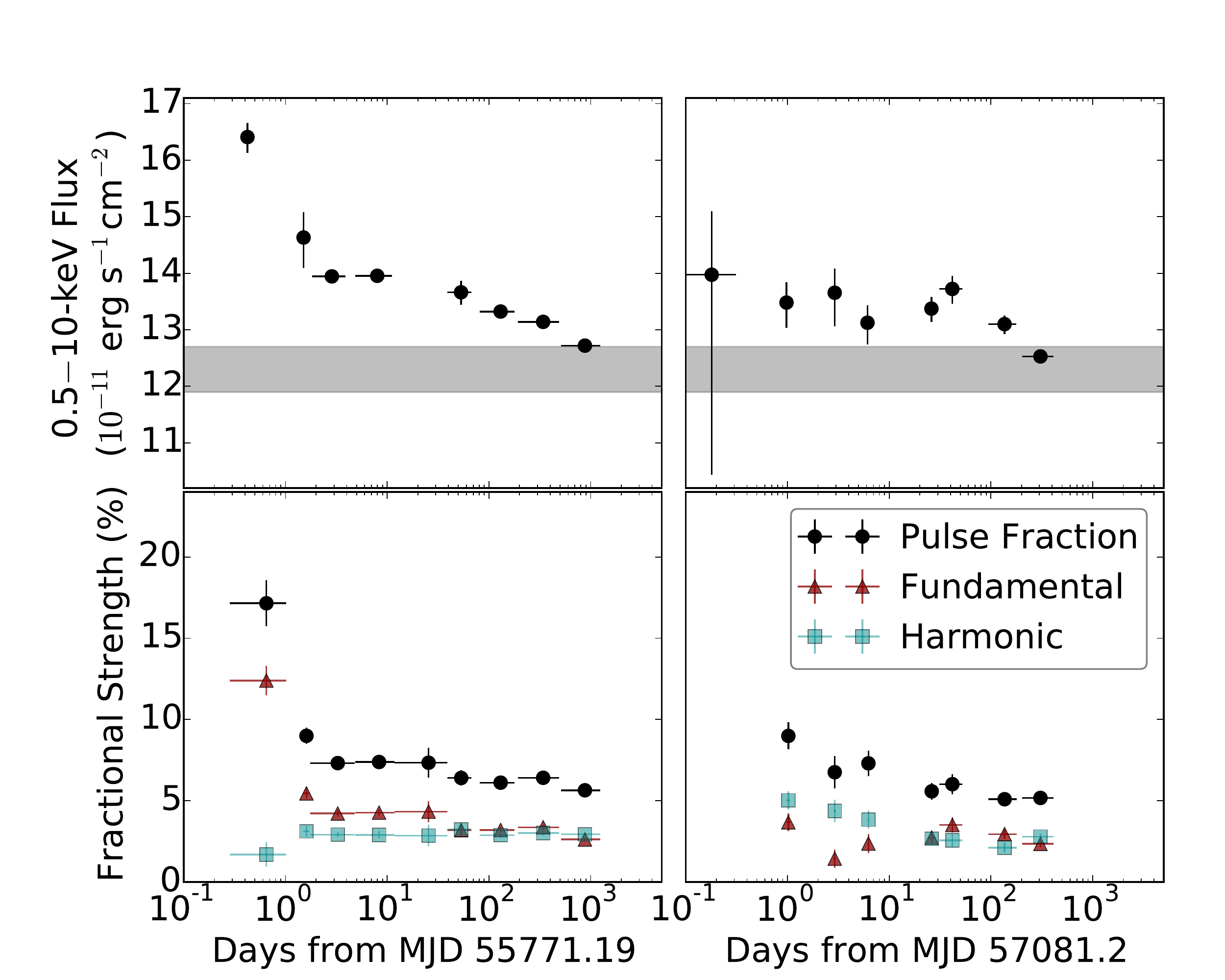}
\caption {Flux \& pulse shape evolution of \source{}. The top panels show the total absorbed 0.5--10-keV flux, binned in equal log-time bins from MJDs 55771.19, and 57081.2. The gray band in the upper panels shows the pre-2011 outburst flux value.
The lower panels show the 0.7--10\,keV RMS pulsed fraction as black circles, the strength in the fundamental (i.e. at the spin frequency) as red triangles, and first harmonic (i.e. at twice the spin frequency), as blue squares, with the same time binning.}
\label{fig:pf}
\end{figure}

We also fitted the long-term decay in X-ray flux following the 2015 glitch.
In this case we required only a single exponential decay, again fixing the quiescent flux at $12.4(4)\times 10^{-11}$ erg s$^{-1}$ cm$^{-2}$.
The best-fit parameters give a flux increase of $F=1.3(3)\times10^{-11}$ erg s$^{-1}$ cm$^{-2}$ with a decay time of $\tau = 160(70)\;$ days with $\chi^2/dof=13.7/21$.
The flux decay can be more clearly seen in Figure~\ref{fig:pf}, where the data have been binned.

The X-ray pulse profile of \source{} has been shown to evolve over a time scale of years \citep{2007ApJ...666...1152D}.
As such, we investigated the evolution of the pulse profile by at the time-scales suggested by the decaying total flux.
To do so, we folded all aforementioned XRT observations into 32-bin profiles using a timing solution presented in Table~\ref{tab:timing}, and created a profile for each equally logarithmically spaced segment of the time series following each of the two timing events.
We then transformed these profiles into their respective Fourier representations, allowing us to quantify both the pulse shape, and root mean squared (RMS) pulsed fraction.
For details on the RMS pulsed fraction, see e.g. \cite{2015ApJ...807...93A}.
In Figure~\ref{fig:pf} we present the pulsed fraction, and power in the first two Fourier harmonics over time.
For reference, in the top panel we show the total absorbed 0.5--10-keV flux binned at the same time resolution.
The pulse profiles themselves can be seen in Figure~\ref{fig:profs}.
It is clear that the pulsed fraction of the source decreased rapidly starting from the 2011 glitch, mirroring the behavior of the total X-ray flux.
This was driven primarily by a decrease in the strength of the fundamental.
Following both glitches the pulse profile shows variability for approximately ten days following the glitch before reattaining the normal state with both the fundamental and harmonic having equal power, as can be seen in Figure~\ref{fig:pf}.

\subsection{ Short-term Flux Enhancement \& Bursts}
\label{sec:shortflux}
\begin{figure}
\centering
\includegraphics[width=\columnwidth]{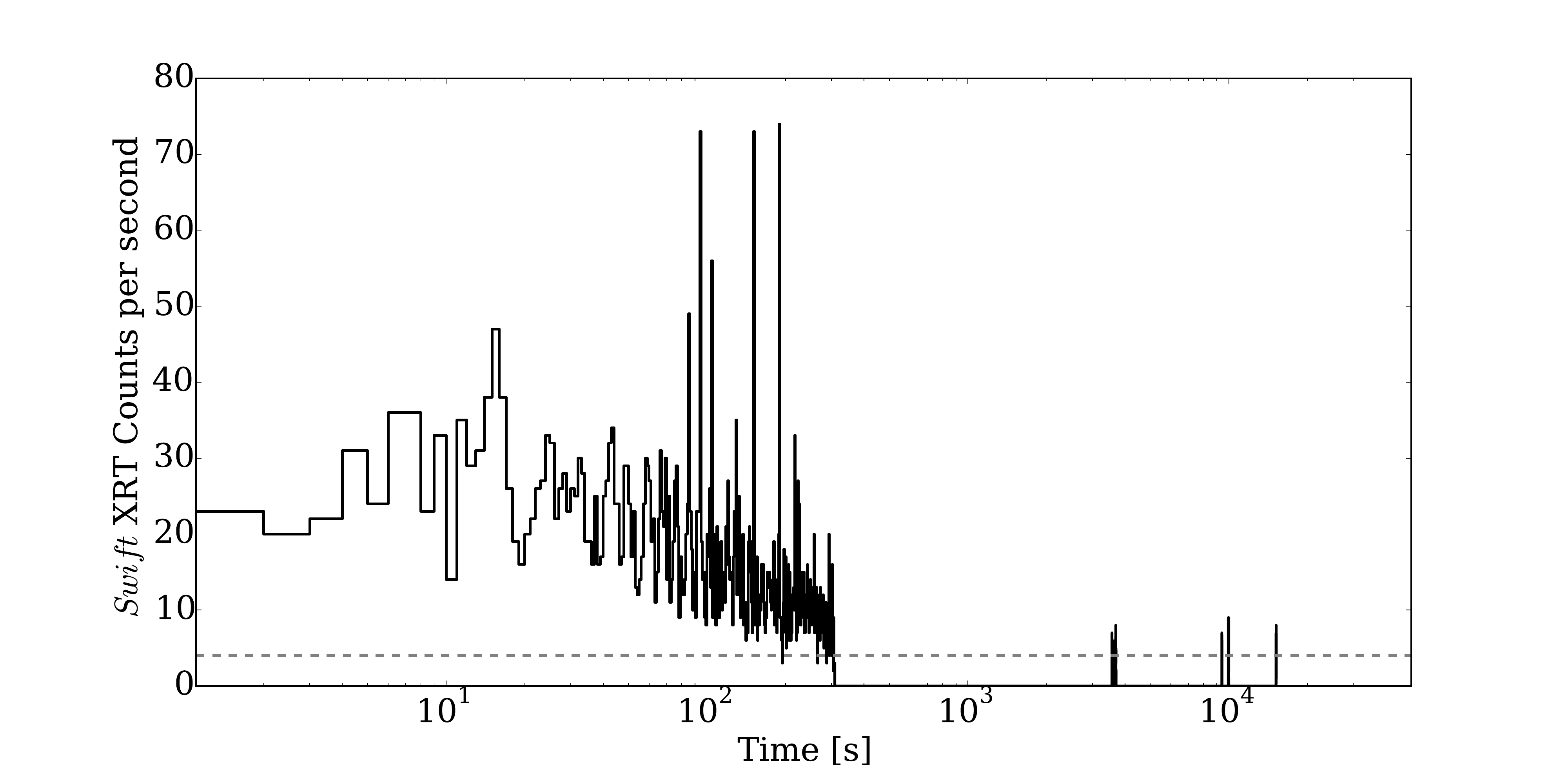}
\caption {{\it Swift} XRT light-curve in the 0.7--10\,keV band for the 2015 Febuary 28 burst observation. The light curve has been binned at 1$\;$s. The gray dashed line represents the average count rate of \source{}.}
\label{fig:lc}
\end{figure}

\begin{table}
\begin{center}
\caption{{\it Swift} XRT-detected burst parameters from \source{} during the 2015 February event.}
\label{tab:bursts}
\begin{tabular}{c c c c}

\hline
\hline
Burst time  & $T_{90}$ & Fluence & Phase \\
s from start & ms & 1--10$\;$keV Photons &$\phi$(0-1) \\
\hline
37.46 & $103\pm5$ & $14\pm4$ & 0.48 \\
66.48 & $ 39\pm14$ & $12\pm3$ & 0.82 \\
85.60 & $110\pm80$ & $19\pm4$ & 0.02 \\
94.80 & $450\pm60$ & $61\pm8$ & 0.08 \\
104.16 & $170\pm50$ & $37\pm6$ & 0.16 \\
151.81 & $260\pm20$ & $56\pm7$ & 0.64 \\
189.28 & $110\pm30$ & $48\pm7$ & 0.95 \\
217.46 & $ 80\pm50$ & $22\pm5$ & 0.19 \\
225.50 & $400\pm300$ & $17\pm4$ & 0.12 \\
256.97 & $500\pm100$ & $19\pm4$ & 0.74 \\
301.77 & $400\pm300$ & $ 7\pm3$ & 0.90 \\
\hline
85 866.90& $30\pm20$ & $ 7\pm3$ & 0.17 \\
\hline
\hline
\end{tabular}
\end{center}
\end{table}
Immediately following the BAT trigger on MJD 57081, 2015 Febuary 28 \citep{2015GCN..17507...1B}, the XRT slewed to observe \source{}.
For the first snapshot of observation, the source was significantly brighter, and harder than in the surrounding observations.
Fitting an absorbed blackbody plus power law was not warranted, as the spectrum is well fit (C-stat of 904 for 947 degrees of freedom) by an absorbed power law with $\Gamma=1.68(5)$ and an absorbed 0.5--10-keV flux of $1.06(3)\times 10^{-9}\,$erg s$^{-1}$ cm$^{-2}$.
However, we note this is an average flux for this snapshot.
The flux evolved on a faster timescale than this, and a more complicated short-term flux evolution is clear.
This is evident in the binned light-curve of the XRT snapshot taken after the BAT trigger, shown in Figure~\ref{fig:lc} with a reference line plotted at the average count rate for \source{}.

We searched all {\it Swift} XRT observations for magnetar-like bursts at timescales of 1, 0.1, and 0.01$\;$s.
This was done following the method of \cite{2011ApJ...739...94S} wherein for each Good Time Interval (GTI), statistically significant deviations from the mean count rate, assuming Poisson statistics, are flagged.
Due to the aforementioned instrumental bursts, the burst search was run only on photons having energies greater than 0.7\,keV.
We detected significant bursts only in the XRT data in the observations following the 2015 February 28 BAT detection \citep{2015GCN..17507...1B}.
Eleven bursts were found in this observation, superimposed on an overall flux decrease with a decay rate on the order $100\,$s.

As well, one burst was detected in the observation one day later.
In Table~\ref{tab:bursts} we present the burst times in seconds from MJD 57081.2043746, the total fluences, $T_{90}$ (the time period which contains 90\% of a bursts fluence) and the phase at which the peak of the burst occurred.
The properties of these bursts are typical for magnetars \citep{2015ApJS..218...11C}.

We also note a third BAT detection of \source{} on  2012 Jan 12 \citep{2012GCN..12829...1B}.
This burst was unaccompanied by any timing event, or long-term radiative change.
While this lack of associated timing event is unique among the three BAT detections presented here, this is not atypical behavior for magnetars.
For example 1E\,1841$-$045 has triggered BAT on several occasions with no measurable change in its timing properties \citep{2015ApJ...807...93A}.

\section{Discussion}
\label{sec:discussion}

In this work, we have presented the evolution of the pulse profile, flux, and timing parameters of \source{} from July 2011 to June 2016.
Over this monitoring campaign, we report two timing events, both accompanied by changes in the radiative behavior of the magnetar, as well as a BAT detected burst with no other associated behavior.

The first timing event occurred in 2011, and was characterized by a simple step-like glitch in spin frequency heralded by a BAT detected bright X-ray burst.
This glitch was followed by an X-ray flux decay with two timescales -- an initial fast decay with $\tau=0.6(2)$\,days and a longer $\tau=510(80)$\,day decay.

The second timing event in the {\it Swift} campaign occurred in 2015.
This glitch had a more complex timing structure requiring a two-component model with a 57(9) day timescale resulting in a net spin down of the pulsar, see Table~\ref{tab:timing}.
This event was also accompanied by a BAT detected bright X-ray burst, as well as
a long lived flux decay with $\tau=160(70)$\,days.
This second event bears a striking resemblance to the 2006 outburst from this same source \citep{2011ApJ...736..138G}.
Both began with a $\sim100\,$s timescale X-ray flux increase, accompanied by several, clustered magnetar-like short X-ray bursts and a timing event resulting in a net spin-down of the pulsar.
The long-term X-ray flux evolution of \source\ is quite typical for a magnetar: the spectrum becoming harder at the time of the outburst; see \S~\ref{sec:shortflux}.

\subsection{Radiative Evolution}

There are two main classes of models to explain the flux evolution of magnetar outbursts: neutron star cooling and magnetospheric relaxation.
Both sets of models can explain the differing timescales for outbursts from the same source, a common feature in magnetar outbursts \citep[e.g.][]{2011A&A...529A..19B, 2004ApJ...609L..67G, 2012ApJ...757...68A}.

In neutron star cooling models, the long-term flux evolution following a magnetar outburst is a thermal relaxation of the crust of the neutron star until it reattains thermal equilibrium with the core.
In this model, the differing timescales for different events in the same source may be related to energy being injected at different depths in the crust of the neutron star \citep{2009ApJ...698.1020B,  2012ApJ...761...66S, 2015ApJ...809L..31D}, with longer timescales corresponding to deeper layers in the crust.

The other major class of models which describe the flux relaxation of magnetars following outbursts involves the untwisting of the magnetosphere \citep[e.g][]{2009ApJ...703.1044B}. 
In this model following the outbursts current carrying field lines, $j$-bundles, have been twisted by crustal motion.
These twists are anchored to the surface and bombard the surface with accelerated charged particles, heating the surface.
As the twist dissipates, the area affected by the bombardment shrinks, and cools.
 This has an expected decay time of $t_{ev}\sim 10^{7}\,\mu_{32}\,\Phi_{10}^{-1} \,A_{11.5}$\,s where $\mu_{32}$ is the magnetic moment in units of $10^{32}$\,G\,cm$^3$, $\Phi_{10}$ is the electric voltage sustaining electron-positron discharge in the magnetosphere in units of $10^{10}$\,V and $A_{11.5}$ the area of the $j$-bundle footprint in units of $10^{11.5}$\,cm$^{2}$.
 With the parameters of \source{}, this timescale should be $t_{ev}\sim10^7\Phi_{10}^{-1}$\,s$\sim150\Phi_{10}^{-1}$\,days, roughly consistent with the flux decay timescale of the 2015 outburst.
In this model, these different timescales would be due to different values of $\Phi_{10}$, as  the magnetic moment of the neutron star has not changed, and the blackbody area has  only changed at the $\sim10\%$ level over this campaign.

\begin{table*}[]
    \footnotesize
    \hspace{-1.5cm}
      \begin{threeparttable}
    \caption{Reported net spin-down glitches.}
    \label{tab:anti}

    \begin{tabular}{ l c c c c l l }

    \hline
    \hline
    Pulsar & B-field & Long-term $\Delta\nu$ & Timescale &Q &Radiative Behavior & Reference\\
     & $10^{14}\,$G & Hz &Days& & & \\
    \hline
    SGR$\,$1900+14 &7.0 & $-1.10(3)\times10^{-4}$&$<$80 & --&giant flare, bursts & \cite{1999ApJ...524L..55W}\\
    PSR$\,$J1846--0258 & 0.49&$-9.52(9)\times10^{-5}$ &127(5)&8.7(2.5) & bursts, flux increase & \cite{2010ApJ...710.1710L}\tnote{1}\\
    1E$\,$2259+586 & 0.59&$-4.5(6)\times10^{-8}$ &$<$4 &--& burst, flux increase& \cite{2013Natur.497..591A}\tnote{2} \\
    1E$\,$2259+586 & 0.59&$-1.2(3)\times10^{-8}$ &--&--& flux increase& \cite{2012MNRAS.419.3109I, 2014ApJ...784...37D} \\
    4U$\,$0142+61 & 1.3&$-1.27(17)\times10^{-8}$ &17(1.7) &1.07(2)& bursts, flux increase & \cite{2011ApJ...736..138G}\tnote{3} \\
    4U$\,$0142+61 & 1.3 &$-3.7(1)\times10^{-8}$ &57(9) &3.6(4)& bursts, flux increase & This work\tnote{4}\\
    $^\dagger$1E$\,$1841--045 &7.0 &$-4.9(6)\times10^{-8}$ &-- & --&-- & \cite{2014MNRAS.440.2916S}\tnote{5} \\
    $^\dagger$XTE$\,$J1810--197 & 2.1&$-7(6)\times10^{-5}$ &51(21) &$\sim$1 &--& \cite{2016MNRAS.458.2088P}\tnote{6} \\
    \hline
    \hline
        \end{tabular}

        \begin{tablenotes}
      \item[$^\dagger$]{Disputed events, see the notes below.}
      \item[1]{Accompanied by decaying spin-up glitch with $\Delta\nu=10.8(4)\times10^{-5}$\,Hz.}
      \item[2]{Possibly accompanied by second glitch event.}
      \item[3]{Accompanied by decaying spin-up glitch with $\Delta\nu=2.0(4)\times10^{-7}$\,Hz.}
      \item[4]{Accompanied by decaying spin-up glitch with $\Delta\nu=5.1(5)\times10^{-8}$\,Hz.}
      \item[5]{Not seen in analysis of same data by \cite{2014ApJ...784...37D}.}
      \item[6]{Glitch reported to completely recover, also not required in analysis of same data by \cite{2016ApJ...820..110C}.\newline}
      \end{tablenotes}
      \end{threeparttable}
\end{table*}

\subsection{Glitches in High-B pulsars}

Glitches in pulsars often result in both a permanent step in spin frequency and components which decay -- typically with the functional form of an exponential.
In standard pulsar glitch theory, glitches and their recoveries are an observable of the coupling between the neutron superfluid and non-superfluid components of the crust \cite[e.g.][]{1975Natur.256...25A}.
For typical glitches, which result in a spin-up of the neutron star, the sudden spin-up is explained as a transfer of angular momentum from the superfluid component to the solid components of the crust.
This will result in a spin-up of the neutron star as the superfluidic component has more angular momentum than the crust due to the crust suffering external spin-down torque.
When the superfluid and solid components become coupled, the superfluid will have a higher angular velocity, and transfer some of this extra angular momentum to the solid crust.
For a more thorough review of pulsar glitch theory, see \cite{2015IJMPD..2430008H}.

In this standard picture, spin-down glitches can occur via the transfer of angular momentum from  the inner crust, which may have slower spinning regions due to plastic crustal deformation caused by the extreme magnetic field \cite[e.g.][]{2000ThompsonSpinDown}.
The long-term X-ray pulse profile and pulsed fraction evolution, such as that presented in \S\ref{sec:longflux},  could be evidence for the slow, plastic deformation of the crust needed to produce slower regions of superfluid \citep{2004ApJ...605..378W, 2002ApJ...574..332T}.

Alternatively, regions of the outer core could be involved due to the strong magnetic fields.
Magnetar strength fields provide a strong pinning of vortices to flux tubes in the outer core, which leads to a rotation lag between the normal and superfluid components which, when relaxed, would lead to a spin-down event \citep{2014ApJ...797L...4K}.

In typical rotation-powered pulsars, the believed internal nature of these spin-up glitches is argued to be indicated by the lack of observed radiative changes. 
This `radiatively quiet' condition does not hold for many glitches in magnetars.
Many magnetar glitches are accompanied by drastic changed in their X-ray flux and pulse profiles \citep[e.g.][]{2003ApJ...588L..93K, 2009ApJ...702..614D}.
This could indicate that external forces are at play in radiatively loud glitches, and these external forces may lead to the net spin-down events reported in several magnetars, including the new event reported here.

We have compiled a list of all published confirmed and candidate net spin-down glitches in Table~\ref{tab:anti}.
In this list, we present the magnitude of the change in spin frequency, $\Delta\nu$.
For net spin-down glitches that are due to over-recovering spin-up glitches, the $\tau_d$ of the glitch is presented.
For events where there is no sign of a recovery, we show an upper limit for the time during which the spin-down had to occur.
We also note that six of the eight reported net spin-down glitches are accompanied by radiative changes\footnote{This rises to six of six if we ignore the two disputed events; see Table~\ref{tab:anti}.}.
This is a much higher fraction of radiatively activity than seen in spin-up events, where only five of twenty-two timing anomalies were accompanied by radiative changes \citep{2014ApJ...784...37D}.

For the net spin-down glitches in which the net spin down is measured to be due to an over-recovery, we also report the recovery fraction, $Q$.
Recovering glitches are often classified by their recovery fraction, defined as $Q\equiv\Delta\nu_d/(\Delta\nu_d+\Delta\nu)$.
In younger pulsars, when recoveries are observed, $Q$ tends to be larger than in older pulsars \citep{2000MNRAS.317..843W}.
For the new glitch we report here,  we measure $Q= 3.6\pm0.4$.
Glitches with $Q>1$ have been seen only twice before:
once following the magnetar-like outburst of the high magnetic field rotation powered pulsar PSR$\;$J1846$-$0258 which had $Q=8.7\pm2.5$ \citep{2010ApJ...710.1710L}, and once during the 2006 outburst of \source{} with $Q=1.07\pm0.02$ \citep{2011ApJ...736..138G}.
Interestingly, all three of these over-recovering glitches were accompanied by radiative behavior.
Indeed, it should be noted that due to the necessary non-continuous monitoring strategies used to study magnetars, all of the net spin-down events, i.e. the `anti-glitches', may be unresolved over-recovering glitches, albeit with a short time scale.
The most constraining limit placed on an unresolved over-recovery time scale is less than four days in the 2012 glitch of the magnetar 1E\,2259$+$586. \citep{2013Natur.497..591A}.

The radiatively loud nature of these spin-down glitches could be symptomatic of external changes, e.g. the magnetosphere \citep{2009ApJ...703.1044B, 2012ApJ...754L..12P, 2013ApJ...774...92P}.
However, magnetars also exhibit typical spin-up glitches that are accompanied by similar radiative changes \citep[e.g.][]{2003ApJ...588L..93K, 2009ApJ...702..614D}, which could indicate that another variable is at play.
One possibility for this is the location of the twist in the magnetosphere.
The observed radiative properties of magnetars are determined by their closed field, whereas the spin-down properties are dominated by the open field lines \citep[e.g.][]{2007ApJ...657..967B, 2009ApJ...703.1044B}.
Therefore, depending on whether or not the closed field line region is affected during a twist event could determine if a glitch is accompanied by a radiative outburst.

\section{Conclusion}
We have presented the results of a five-year monitoring campaign of \source{} using the {\it Swift} XRT.
Over this campaign,  we have shown that \source\ has had two X-ray outbursts associated with timing events, one in 2011, and a second in 2015.
The 2011 outburst was accompanied by a simple step-like spin-up glitch in spin frequency \citep{2014ApJ...784...37D}.
The 2015 outburst was accompanied by an unusual glitch, starting with a spin-up in the spin frequency which decayed with a 57(9) day timescale, resulting in a net spin down of the pulsar.

As the sample of timing anomalies in magnetars continues to grow, we are detecting more net spin-down events, which have not been seen in the normal radio pulsar population.
This strongly implicates the influence of a large magnetic field in spin-down events and, coupled with the radiatively loud nature of the plurality of spin-down events, suggests an origin in the magnetosphere of the star.

\smallskip
\noindent {\it Acknowledgements:} We are grateful to the {\it Swift} team for their flexibility in the scheduling of the timing monitoring campaign of \source{}.
We thank A. Cumming for helpful discussions.
RFA acknowledges support from an  NSERC  Alexander  Graham  Bell  Canada  Graduate  Scholarship.
VMK receives support from an NSERC Discovery Grant and Accelerator Supplement, Centre de Recherche en Astrophysique du Quebec, an R. Howard Webster Foundation Fellowship from the Canadian Institute for Advanced Study, the Canada Research Chairs Program and the Lorne Trottier Chair in Astrophysics and Cosmology.
PS acknowledges support from a Schulich Graduate Fellowship from McGill University.
We acknowledge the use of public data from the {\it Swift} data archive.
This research has made use of data obtained through the High Energy Astrophysics Science Archive Research Center Online Service, provided by the NASA/Goddard Space Flight Center.

\bibliography{/homes/borgii/rarchiba/Papers/rfa}{}

\begin{thebibliography}{}
\providecommand\natexlab[1]{#1}
\providecommand\JournalTitle[1]{#1}

\bibitem[{{An} {et~al.}(2012){An}, {Kaspi}, {Tomsick}, {Cumming}, {Bodaghee},
  {Gotthelf}, \& {Rahoui}}]{2012ApJ...757...68A}
{An}, H., {Kaspi}, V.~M., {Tomsick}, J.~A., {et~al.} 2012,
  \href{http://dx.doi.org/10.1088/0004-637X/757/1/68}{\JournalTitle{\apj}, 757,
  68}

\bibitem[{{An} {et~al.}(2015){An}, {Archibald}, {Hasco{\"e}t}, {Kaspi},
  {Beloborodov}, {Archibald}, {Beardmore}, {Boggs}, {Christensen}, {Craig},
  {Gehrels}, {Hailey}, {Harrison}, {Kennea}, {Kouveliotou}, {Stern}, {Younes},
  \& {Zhang}}]{2015ApJ...807...93A}
{An}, H., {Archibald}, R.~F., {Hasco{\"e}t}, R., {et~al.} 2015,
  \href{http://dx.doi.org/10.1088/0004-637X/807/1/93}{\JournalTitle{\apj}, 807,
  93}

\bibitem[{{Anderson} \& {Itoh}(1975)}]{1975Natur.256...25A}
{Anderson}, P.~W., \& {Itoh}, N. 1975,
  \href{http://dx.doi.org/10.1038/256025a0}{\JournalTitle{\nat}, 256, 25}

\bibitem[{{Archibald} {et~al.}(2015){Archibald}, {Kaspi}, {Ng}, {Scholz},
  {Beardmore}, {Gehrels}, \& {Kennea}}]{2015ApJ...800...33A}
{Archibald}, R.~F., {Kaspi}, V.~M., {Ng}, C.-Y., {et~al.} 2015,
  \href{http://dx.doi.org/10.1088/0004-637X/800/1/33}{\JournalTitle{\apj}, 800,
  33}

\bibitem[{{Archibald} {et~al.}(2013){Archibald}, {Kaspi}, {Ng},
  {Gourgouliatos}, {Tsang}, {Scholz}, {Beardmore}, {Gehrels}, \&
  {Kennea}}]{2013Natur.497..591A}
---. 2013, \href{http://dx.doi.org/10.1038/nature12159}{\JournalTitle{\nat},
  497, 591}

\bibitem[{{Barthelmy} {et~al.}(2015){Barthelmy}, {Gehrels}, {Kennea}, {Lien},
  {Marshall}, {Maselli}, {Palmer}, \& {Siegel}}]{2015GCN..17507...1B}
{Barthelmy}, S.~D., {Gehrels}, N., {Kennea}, J.~A., {et~al.} 2015,
  \JournalTitle{GRB Coordinates Network}, 17507, 1

\bibitem[{{Barthelmy} {et~al.}(2012){Barthelmy}, {D'Elia}, {Gehrels}, {Gendre},
  {Guidorzi}, {Krimm}, {Lien}, {Littlejohns}, {Palmer}, {Romano}, {Sbarufatti},
  {Starling}, {Stratta}, \& {Tagliaferri}}]{2012GCN..12829...1B}
{Barthelmy}, S.~D., {D'Elia}, V., {Gehrels}, N., {et~al.} 2012,
  \JournalTitle{GRB Coordinates Network}, 12829

\bibitem[{{Beloborodov}(2009)}]{2009ApJ...703.1044B}
{Beloborodov}, A.~M. 2009,
  \href{http://dx.doi.org/10.1088/0004-637X/703/1/1044}{\JournalTitle{\apj},
  703, 1044}

\bibitem[{{Beloborodov} \& {Thompson}(2007)}]{2007ApJ...657..967B}
{Beloborodov}, A.~M., \& {Thompson}, C. 2007,
  \href{http://dx.doi.org/10.1086/508917}{\JournalTitle{\apj}, 657, 967}

\bibitem[{{Bernardini} {et~al.}(2011){Bernardini}, {Israel}, {Stella},
  {Turolla}, {Esposito}, {Rea}, {Zane}, {Tiengo}, {Campana}, {G{\"o}tz},
  {Mereghetti}, \& {Romano}}]{2011A&A...529A..19B}
{Bernardini}, F., {Israel}, G.~L., {Stella}, L., {et~al.} 2011,
  \href{http://dx.doi.org/10.1051/0004-6361/201016197}{\JournalTitle{\aap},
  529, A19}

\bibitem[{{Brown} \& {Cumming}(2009)}]{2009ApJ...698.1020B}
{Brown}, E.~F., \& {Cumming}, A. 2009,
  \href{http://dx.doi.org/10.1088/0004-637X/698/2/1020}{\JournalTitle{\apj},
  698, 1020}

\bibitem[{{Burrows} {et~al.}(2005){Burrows}, {Hill}, {Nousek}, {Kennea},
  {Wells}, {Osborne}, {Abbey}, {Beardmore}, {Mukerjee}, {Short}, {Chincarini},
  {Campana}, {Citterio}, {Moretti}, {Pagani}, {Tagliaferri}, {Giommi},
  {Capalbi}, {Tamburelli}, {Angelini}, {Cusumano}, {Br{\"a}uninger}, {Burkert},
  \& {Hartner}}]{2005SSRv..120..165B}
{Burrows}, D.~N., {Hill}, J.~E., {Nousek}, J.~A., {et~al.} 2005,
  \href{http://dx.doi.org/10.1007/s11214-005-5097-2}{\JournalTitle{\ssr}, 120,
  165}

\bibitem[{{{\c S}a{\c s}maz Mu{\c s}} {et~al.}(2014){{\c S}a{\c s}maz Mu{\c
  s}}, {Ayd{\i}n}, \& {G{\"o}{\u g}{\"u}{\c s}}}]{2014MNRAS.440.2916S}
{{\c S}a{\c s}maz Mu{\c s}}, S., {Ayd{\i}n}, B., \& {G{\"o}{\u g}{\"u}{\c s}},
  E. 2014, \href{http://dx.doi.org/10.1093/mnras/stu436}{\JournalTitle{\mnras},
  440, 2916}

\bibitem[{{Camilo} {et~al.}(2016){Camilo}, {Ransom}, {Halpern}, {Alford},
  {Cognard}, {Reynolds}, {Johnston}, {Sarkissian}, \& {van
  Straten}}]{2016ApJ...820..110C}
{Camilo}, F., {Ransom}, S.~M., {Halpern}, J.~P., {et~al.} 2016,
  \href{http://dx.doi.org/10.3847/0004-637X/820/2/110}{\JournalTitle{\apj},
  820, 110}

\bibitem[{{Cash}(1979)}]{1979ApJ...228..939C}
{Cash}, W. 1979, \href{http://dx.doi.org/10.1086/156922}{\JournalTitle{\apj},
  228, 939}

\bibitem[{{Collazzi} {et~al.}(2015){Collazzi}, {Kouveliotou}, {van der Horst},
  {Younes}, {Kaneko}, {G{\"o}{\u g}{\"u}{\c s}}, {Lin}, {Granot}, {Finger},
  {Chaplin}, {Huppenkothen}, {Watts}, {von Kienlin}, {Baring}, {Gruber},
  {Bhat}, {Gibby}, {Gehrels}, {McEnery}, {van der Klis}, \&
  {Wijers}}]{2015ApJS..218...11C}
{Collazzi}, A.~C., {Kouveliotou}, C., {van der Horst}, A.~J., {et~al.} 2015,
  \href{http://dx.doi.org/10.1088/0067-0049/218/1/11}{\JournalTitle{\apjs},
  218, 11}

\bibitem[{{Deibel} {et~al.}(2015){Deibel}, {Cumming}, {Brown}, \&
  {Page}}]{2015ApJ...809L..31D}
{Deibel}, A., {Cumming}, A., {Brown}, E.~F., \& {Page}, D. 2015,
  \href{http://dx.doi.org/10.1088/2041-8205/809/2/L31}{\JournalTitle{\apjl},
  809, L31}

\bibitem[{{Dib} \& {Kaspi}(2014)}]{2014ApJ...784...37D}
{Dib}, R., \& {Kaspi}, V.~M. 2014,
  \href{http://dx.doi.org/10.1088/0004-637X/784/1/37}{\JournalTitle{\apj}, 784,
  37}

\bibitem[{Dib {et~al.}(2007)Dib, Kaspi, \& Gavriil}]{2007ApJ...666...1152D}
Dib, R., Kaspi, V.~M., \& Gavriil, F.~P. 2007,
  \href{http://stacks.iop.org/0004-637X/666/i=2/a=1152}{\JournalTitle{\apj},
  666, 1152}

\bibitem[{{Dib} {et~al.}(2008){Dib}, {Kaspi}, \&
  {Gavriil}}]{2008APJ.673..1044D}
{Dib}, R., {Kaspi}, V.~M., \& {Gavriil}, F.~P. 2008,
  \href{http://dx.doi.org/10.1086/524653}{\JournalTitle{\apj}, 673, 1044}

\bibitem[{{Dib} {et~al.}(2009){Dib}, {Kaspi}, \&
  {Gavriil}}]{2009ApJ...702..614D}
---. 2009,
  \href{http://dx.doi.org/10.1088/0004-637X/702/1/614}{\JournalTitle{\apj},
  702, 614}

\bibitem[{{Dierckx}(1975)}]{splineref}
{Dierckx}, P. 1975, \JournalTitle{J.Comp.Appl.Maths}, 1, 165

\bibitem[{{Gavriil} {et~al.}(2011{\natexlab{a}}){Gavriil}, {Dib}, \&
  {Kaspi}}]{2011ApJ...736..138G}
{Gavriil}, F.~P., {Dib}, R., \& {Kaspi}, V.~M. 2011{\natexlab{a}},
  \href{http://dx.doi.org/10.1088/0004-637X/736/2/138}{\JournalTitle{\apj},
  736, 138}

\bibitem[{{Gavriil} \& {Kaspi}(2004)}]{2004ApJ...609L..67G}
{Gavriil}, F.~P., \& {Kaspi}, V.~M. 2004,
  \href{http://dx.doi.org/10.1086/422751}{\JournalTitle{\apjl}, 609, L67}

\bibitem[{{Gavriil} {et~al.}(2011{\natexlab{b}}){Gavriil}, {Kaspi},
  {Livingstone}, {Scholz}, \& {Archibald}}]{2011ATel.3520....1G}
{Gavriil}, F.~P., {Kaspi}, V.~M., {Livingstone}, M.~A., {Scholz}, P., \&
  {Archibald}, R. 2011{\natexlab{b}}, \JournalTitle{The Astronomer's Telegram},
  3520, 1

\bibitem[{{Giacconi} {et~al.}(1972){Giacconi}, {Murray}, {Gursky}, {Kellogg},
  {Schreier}, \& {Tananbaum}}]{1972ApJ...178..281G}
{Giacconi}, R., {Murray}, S., {Gursky}, H., {et~al.} 1972,
  \href{http://dx.doi.org/10.1086/151790}{\JournalTitle{\apj}, 178, 281}

\bibitem[{{Haskell} \& {Melatos}(2015)}]{2015IJMPD..2430008H}
{Haskell}, B., \& {Melatos}, A. 2015,
  \href{http://dx.doi.org/10.1142/S0218271815300086}{\JournalTitle{International
  Journal of Modern Physics D}, 24, 1530008}

\bibitem[{{Hobbs} {et~al.}(2006){Hobbs}, {Edwards}, \&
  {Manchester}}]{2006MNRAS.369..655H}
{Hobbs}, G.~B., {Edwards}, R.~T., \& {Manchester}, R.~N. 2006,
  \href{http://dx.doi.org/10.1111/j.1365-2966.2006.10302.x}{\JournalTitle{\mnras},
  369, 655}

\bibitem[{{Hulleman} {et~al.}(2004){Hulleman}, {van Kerkwijk}, \&
  {Kulkarni}}]{2004A&A...416.1037H}
{Hulleman}, F., {van Kerkwijk}, M.~H., \& {Kulkarni}, S.~R. 2004,
  \href{http://dx.doi.org/10.1051/0004-6361:20031756}{\JournalTitle{\aap}, 416,
  1037}

\bibitem[{{I{\c c}dem} {et~al.}(2012){I{\c c}dem}, {Baykal}, \&
  {Inam}}]{2012MNRAS.419.3109I}
{I{\c c}dem}, B., {Baykal}, A., \& {Inam}, S.~{\c C}. 2012,
  \href{http://dx.doi.org/10.1111/j.1365-2966.2011.19953.x}{\JournalTitle{\mnras},
  419, 3109}

\bibitem[{{Israel} {et~al.}(1994){Israel}, {Mereghetti}, \&
  {Stella}}]{1994ApJ...433L..25I}
{Israel}, G.~L., {Mereghetti}, S., \& {Stella}, L. 1994,
  \href{http://dx.doi.org/10.1086/187539}{\JournalTitle{\apjl}, 433, L25}

\bibitem[{{Kantor} \& {Gusakov}(2014)}]{2014ApJ...797L...4K}
{Kantor}, E.~M., \& {Gusakov}, M.~E. 2014,
  \href{http://dx.doi.org/10.1088/2041-8205/797/1/L4}{\JournalTitle{\apjl},
  797, L4}

\bibitem[{{Kaspi} {et~al.}(2003){Kaspi}, {Gavriil}, {Woods}, {Jensen},
  {Roberts}, \& {Chakrabarty}}]{2003ApJ...588L..93K}
{Kaspi}, V.~M., {Gavriil}, F.~P., {Woods}, P.~M., {et~al.} 2003,
  \href{http://dx.doi.org/10.1086/375683}{\JournalTitle{\apjl}, 588, L93}

\bibitem[{{Kaspi} {et~al.}(2000){Kaspi}, {Lackey}, \&
  {Chakrabarty}}]{2000ApJ...537L..31K}
{Kaspi}, V.~M., {Lackey}, J.~R., \& {Chakrabarty}, D. 2000,
  \href{http://dx.doi.org/10.1086/312758}{\JournalTitle{\apjl}, 537, L31}

\bibitem[{{Livingstone} {et~al.}(2010){Livingstone}, {Kaspi}, \&
  {Gavriil}}]{2010ApJ...710.1710L}
{Livingstone}, M.~A., {Kaspi}, V.~M., \& {Gavriil}, F.~P. 2010,
  \href{http://dx.doi.org/10.1088/0004-637X/710/2/1710}{\JournalTitle{\apj},
  710, 1710}

\bibitem[{{Livingstone} {et~al.}(2009){Livingstone}, {Ransom}, {Camilo},
  {Kaspi}, {Lyne}, {Kramer}, \& {Stairs}}]{2009LivingstoneTiming}
{Livingstone}, M.~A., {Ransom}, S.~M., {Camilo}, F., {et~al.} 2009,
  \href{http://dx.doi.org/10.1088/0004-637X/706/2/1163}{\JournalTitle{\apj},
  706, 1163}

\bibitem[{{Mereghetti} {et~al.}(2015){Mereghetti}, {Pons}, \&
  {Melatos}}]{2015SSRv..191..315M}
{Mereghetti}, S., {Pons}, J.~A., \& {Melatos}, A. 2015,
  \href{http://dx.doi.org/10.1007/s11214-015-0146-y}{\JournalTitle{\ssr}, 191,
  315}

\bibitem[{{Olausen} \& {Kaspi}(2014)}]{2014ApJS..212....6O}
{Olausen}, S.~A., \& {Kaspi}, V.~M. 2014,
  \href{http://dx.doi.org/10.1088/0067-0049/212/1/6}{\JournalTitle{\apjs}, 212,
  6}

\bibitem[{{Parfrey} {et~al.}(2012){Parfrey}, {Beloborodov}, \&
  {Hui}}]{2012ApJ...754L..12P}
{Parfrey}, K., {Beloborodov}, A.~M., \& {Hui}, L. 2012,
  \href{http://dx.doi.org/10.1088/2041-8205/754/1/L12}{\JournalTitle{\apjl},
  754, L12}

\bibitem[{{Parfrey} {et~al.}(2013){Parfrey}, {Beloborodov}, \&
  {Hui}}]{2013ApJ...774...92P}
---. 2013,
  \href{http://dx.doi.org/10.1088/0004-637X/774/2/92}{\JournalTitle{\apj}, 774,
  92}

\bibitem[{{Pintore} {et~al.}(2016){Pintore}, {Bernardini}, {Mereghetti},
  {Esposito}, {Turolla}, {Rea}, {Coti Zelati}, {Israel}, {Tiengo}, \&
  {Zane}}]{2016MNRAS.458.2088P}
{Pintore}, F., {Bernardini}, F., {Mereghetti}, S., {et~al.} 2016,
  \href{http://dx.doi.org/10.1093/mnras/stw351}{\JournalTitle{\mnras}, 458,
  2088}

\bibitem[{{Rea} \& {Esposito}(2011)}]{2011ASSP...21..247R}
{Rea}, N., \& {Esposito}, P. 2011,
  \href{http://dx.doi.org/10.1007/978-3-642-17251-9_21}{\JournalTitle{Astrophysics
  and Space Science Proceedings}, 21, 247}

\bibitem[{{Rea} {et~al.}(2013){Rea}, {Israel}, {Pons}, {Turolla}, {Vigan{\`o}},
  {Zane}, {Esposito}, {Perna}, {Papitto}, {Terreran}, {Tiengo}, {Salvetti},
  {Girart}, {Palau}, {Possenti}, {Burgay}, {G{\"o}{\u g}{\"u}{\c s}},
  {Caliandro}, {Kouveliotou}, {G{\"o}tz}, {Mignani}, {Ratti}, \&
  {Stella}}]{2013ApJ...770...65R}
{Rea}, N., {Israel}, G.~L., {Pons}, J.~A., {et~al.} 2013,
  \href{http://dx.doi.org/10.1088/0004-637X/770/1/65}{\JournalTitle{\apj}, 770,
  65}

\bibitem[{{Scholz} {et~al.}(2014{\natexlab{a}}){Scholz}, {Archibald}, {Kaspi},
  {Ng}, {Beardmore}, {Gehrels}, \& {Kennea}}]{2014ApJ...783...99S}
{Scholz}, P., {Archibald}, R.~F., {Kaspi}, V.~M., {et~al.} 2014{\natexlab{a}},
  \href{http://dx.doi.org/10.1088/0004-637X/783/2/99}{\JournalTitle{\apj}, 783,
  99}

\bibitem[{{Scholz} \& {Kaspi}(2011)}]{2011ApJ...739...94S}
{Scholz}, P., \& {Kaspi}, V.~M. 2011,
  \href{http://dx.doi.org/10.1088/0004-637X/739/2/94}{\JournalTitle{\apj}, 739,
  94}

\bibitem[{{Scholz} {et~al.}(2014{\natexlab{b}}){Scholz}, {Kaspi}, \&
  {Cumming}}]{2014ApJ...786...62S}
{Scholz}, P., {Kaspi}, V.~M., \& {Cumming}, A. 2014{\natexlab{b}},
  \href{http://dx.doi.org/10.1088/0004-637X/786/1/62}{\JournalTitle{\apj}, 786,
  62}

\bibitem[{{Scholz} {et~al.}(2012){Scholz}, {Ng}, {Livingstone}, {Kaspi},
  {Cumming}, \& {Archibald}}]{2012ApJ...761...66S}
{Scholz}, P., {Ng}, C.-Y., {Livingstone}, M.~A., {et~al.} 2012,
  \href{http://dx.doi.org/10.1088/0004-637X/761/1/66}{\JournalTitle{\apj}, 761,
  66}

\bibitem[{{Thompson} \& {Duncan}(1995)}]{1995MNRAS.275..255T}
{Thompson}, C., \& {Duncan}, R.~C. 1995, \JournalTitle{\mnras}, 275, 255

\bibitem[{{Thompson} \& {Duncan}(1996)}]{1996ApJ...473..322T}
---. 1996, \href{http://dx.doi.org/10.1086/178147}{\JournalTitle{\apj}, 473,
  322}

\bibitem[{{Thompson} {et~al.}(2000){Thompson}, {Duncan}, {Woods},
  {Kouveliotou}, {Finger}, \& {van Paradijs}}]{2000ThompsonSpinDown}
{Thompson}, C., {Duncan}, R.~C., {Woods}, P.~M., {et~al.} 2000,
  \href{http://dx.doi.org/10.1086/317072}{\JournalTitle{\apj}, 543, 340}

\bibitem[{{Thompson} {et~al.}(2002){Thompson}, {Lyutikov}, \&
  {Kulkarni}}]{2002ApJ...574..332T}
{Thompson}, C., {Lyutikov}, M., \& {Kulkarni}, S.~R. 2002,
  \href{http://dx.doi.org/10.1086/340586}{\JournalTitle{\apj}, 574, 332}

\bibitem[{{Verner} {et~al.}(1996){Verner}, {Ferland}, {Korista}, \&
  {Yakovlev}}]{1996ApJ...465..487V}
{Verner}, D.~A., {Ferland}, G.~J., {Korista}, K.~T., \& {Yakovlev}, D.~G. 1996,
  \href{http://dx.doi.org/10.1086/177435}{\JournalTitle{\apj}, 465, 487}

\bibitem[{{Wang} {et~al.}(2000){Wang}, {Manchester}, {Pace}, {Bailes}, {Kaspi},
  {Stappers}, \& {Lyne}}]{2000MNRAS.317..843W}
{Wang}, N., {Manchester}, R.~N., {Pace}, R.~T., {et~al.} 2000,
  \href{http://dx.doi.org/10.1046/j.1365-8711.2000.03713.x}{\JournalTitle{\mnras},
  317, 843}

\bibitem[{{Wilms} {et~al.}(2000){Wilms}, {Allen}, \&
  {McCray}}]{2000ApJ...542..914W}
{Wilms}, J., {Allen}, A., \& {McCray}, R. 2000,
  \href{http://dx.doi.org/10.1086/317016}{\JournalTitle{\apj}, 542, 914}

\bibitem[{{Woods} {et~al.}(1999){Woods}, {Kouveliotou}, {van Paradijs},
  {Finger}, {Thompson}, {Duncan}, {Hurley}, {Strohmayer}, {Swank}, \&
  {Murakami}}]{1999ApJ...524L..55W}
{Woods}, P.~M., {Kouveliotou}, C., {van Paradijs}, J., {et~al.} 1999,
  \href{http://dx.doi.org/10.1086/312297}{\JournalTitle{\apjl}, 524, L55}

\bibitem[{{Woods} {et~al.}(2004){Woods}, {Kaspi}, {Thompson}, {Gavriil},
  {Marshall}, {Chakrabarty}, {Flanagan}, {Heyl}, \&
  {Hernquist}}]{2004ApJ...605..378W}
{Woods}, P.~M., {Kaspi}, V.~M., {Thompson}, C., {et~al.} 2004,
  \href{http://dx.doi.org/10.1086/382233}{\JournalTitle{\apj}, 605, 378}

\end{thebibliography}
\bibliographystyle{yahapj}

\end{document}